\documentclass[usenatbib]{emulateapj}
\slugcomment{{\sc Accepted to ApJL:} March 25, 2008}
\shortauthors{Cowan \& Agol}
\shorttitle{Mapping Exoplanets}
\begin{document}

\title{Inverting Phase Functions to Map Exoplanets}

\author{Nicolas B. Cowan\altaffilmark{1},
	 Eric Agol\altaffilmark{1},
	 }

\altaffiltext{1}{Astronomy Department, University of Washington,
   Box 351580, Seattle, WA  98195}
\email{cowan@astro.washington.edu agol@astro.washington.edu}

\begin{abstract}
We describe how to generate a longitudinal brightness map for a tidally locked exoplanet from its phase function light curve. We operate under a number of simplifying assumptions, neglecting limb darkening/brightening, star spots, detector ramps, as well as time-variability over a single planetary rotation. We develop the transformation from a planetary brightness map to a phase function light curve and simplify the expression for the case of an edge-on system. We introduce two models---composed of longitudinal slices of uniform brightness, and sinusoidally varying maps, respectively---which greatly simplify the transformation from map to light curve. We discuss numerical approaches to extracting a longitudinal map from a phase function light curve, explaining how to estimate the uncertainty in a computed map and how to choose an appropriate number of fit parameters. We demonstrate these techniques on a simulated map and discuss the uses and limitations of longitudinal maps. The sinusoidal model provides a better fit to the planet's underlying brightness map, although the slice model is more appropriate for light curves which only span a fraction of the planet's orbit. Regardless of which model is used, we find that there is a maximum of $\sim5$ free parameters which can be meaningfully fit based on a full phase function light curve, due to the insensitivity of the latter to certain modes of the map. This is sufficient to determine the longitudes of primary equatorial hot-spots and cold-spots, as well as the presence of secondary maxima/minima.  
\end{abstract}

\keywords{\
methods: data analysis ---
(stars:) planetary systems ---
}

\section{Introduction}
Observations of secondary eclipses in exoplanetary systems, starting with HD~209458b \citep{Deming_2005} and TrES-1 \citep{Charbonneau_2005}, have made it possible to measure the integrated day-side flux of hot Jupiters \citep[for a review of transiting exoplanet science see][]{Charbonneau_2007}. By carefully studying of the shape of the ingress and egress of secondary eclipses, it should eventually be possible to map the day-side of such planets \citep{Williams_2006, Rauscher_2007}. To characterize the planet's longitudinal temperature profile at all longitudes, however, observations must be made at a variety of points in the planet's orbit. The first successful observations of this sort were reported by \cite{Harrington_2006}, \cite{Knutson_2007} and \cite{Cowan_2007}. The unprecedented quality of the data in \cite{Knutson_2007} and \cite{Knutson_2008} made it possible to model the planet not merely as day and night hemispheres, but to divide the planet into longitudinal slices, hence producing the first (albeit coarse) maps of an exoplanet under restricted assumptions. 

In this paper we elaborate on the inversion techniques which one can use to obtain a longitudinal brightness map from the light curve of phase variations. Such maps promise to be powerful diagnostic tools for simulations of hot Jupiter atmospheric dynamics because they are nearly model-independent.

\subsection{Simplifying Assumptions}
The detailed study of coupled radiative transfer and dynamics in the atmospheres of hot Jupiters is a tremendously complex science \citep[for a current review of the field, see][]{Showman_2007}. Certain models of hot Jupiter atmospheres predict significant variability in the integrated brightness of the planets\citep[][and references therein]{Rauscher_2008}. Although it may be possible to glean useful information from the phase function light curves of planets with such variable atmospheres, we elect in this Letter to ignore time variability, in line with the detailed modeling of \cite{Cooper_2005}. We also choose to ignore limb darkening/brightening since most atmospheric models of hot Jupiters predict infrared photon absorption lengths much shorter than the scale height of temperature variations. In any case, the presence of limb darkening would not significantly change the present analysis since the limb does not contribute much to the integrated light of the planet (we verify this in \S~3.4).    

Star spots can plague phase function observations since they change the overall brightness of a planet/star system as they rotate into and out of view. Fortunately, star spots vary on longer timescales ($\sim$weeks) than the orbital periods of currently known transiting exoplanets ($\sim$days), and have larger variations in the optical/near-IR which can be used to characterize and subtract the star spot variation \citep[eg:][]{Knutson_2008}. Furthermore, some instruments on the Spitzer Space Telescope \citep{Werner_2004} exhibit detector ramps which distort phase function observations, especially near the beginning of a time series \citep[eg:][]{Knutson_2007}. Both of these observational challenges can be overcome through longer observations: full orbits or more. For the purposes of this Letter, we assume that any---and all---variations in the infrared brightness of an extrasolar planetary system are due to changes in the flux from the planet as different portions of the planet rotate in and out of view. Stellar variability, which contributes to the photometric scatter in the light curve, will likely be the limiting factor in future phase function analysis.   

The mapping technique described here is predicated on the known rotation rate of the planet. As such, it is only applicable to tidally locked planets, where the rotation and orbital periods are identical. Most hot Jupiters are tidally locked so this is not a problematic restriction. Eccentric hot Jupiters are thought to be in pseudo-synchronous rotation, in which case their rotational period can be derived from their orbital parameters. It is unlikely that eccentric planets have steady-state atmospheric dynamics and hence, although the light curve inversion methodology described in this Letter may be applicable, it is not clear that the result would be a ``map''.

\section{From Maps to Light Curves} \label{map_to_lc}
The object of this Letter is to demonstrate how to transform an observed light curve, $F(\xi)$, into a longitudinal brightness map of the planet, where the planet's phase, $\xi$, is the angle---in the plane of the planet's orbit---between the planet, its host star and the planet's position at superior conjunction ($\xi=0$ at superior conjunction; $\xi=\pi$ at inferior conjunction). For edge-on systems, $\xi$ corresponds to the observer--planet--star angle ($\xi = 0$ at secondary eclipse; $\xi=\pi$ at transit). In the interest of simplicity, we ignore the dips in light due to transit and secondary eclipse which occur in transiting systems, although these are crucial in pinning down the absolute flux of the planet (for non-transiting planets $F(\xi)$ is only known to within an additive constant). 

If $\alpha$ and $\beta$ are the longitude and latitude on the planetary disk as seen by an observer and $I$ is the intensity map of the planet, the total flux from the planet as seen by that observer is:
\begin{equation} \label{flux_equation}
F(\xi) = \int_{0}^{\pi}\int_{-\pi/2}^{\pi/2} I(\alpha,\beta) \cos\alpha \sin^{2}\beta d\alpha d\beta.
\end{equation}   
For a tidally locked planet, it is possible to define longitude, $\phi$, and latitude, $\theta$, in the planet's rotating frame, such that $\phi=0$ at the sub-stellar point, $\theta=0$ at the planet's north pole, and $\phi$ increases in the direction of rotation of the planet. The rotation that relates $(\alpha,\beta)$ to $(\phi,\theta)$ can be expressed in terms of Euler angles: $\Phi = \pi/2$, $\Theta=\pi/2-i$, $\Psi = \xi - \pi/2$, where $i$ is the inclination of the planet's orbit ($i = 0$ for a face-on orbit; $i=\frac{\pi}{2}$ for an edge-on orbit). The rotation leads to a system of three coupled equations which can be solved for $\alpha(\phi,\theta)$ and $\beta(\phi,\theta)$. By inserting these expressions into Equation~\ref{flux_equation}, one obtains an equation for the observed flux from the planet as a function of the specific intensity of the planet at different longitudes and latitudes---in effect, a transformation from a brightness map to a light curve: $I(\theta, \phi) \to F(\xi)$.

\subsection{Edge-on Orbits}
Although the formalism above is sufficient to produce a light curve from a two-dimensional planetary map, it is instructive to further constrain the problem by considering planets with inclinations of \emph{precisely} $\pi/2$, with the understanding that the resulting solutions should be approximately correct for transiting planets. In the \emph{worst-case} scenario of a transiting system with $i=80^{\circ}$ and $I(\phi, \theta)$ constant in $\theta$, the edge-on approximation leads to 5\% errors in $F(\xi)$. For edge-on orbits, the rotation relating $(\alpha,\beta)$ to $(\phi,\theta)$ simplifies to $\alpha = \phi + \xi$ and $\beta = \theta$. The expression for $F(\xi)$ can then be written as
 \begin{equation}
F(\xi) = \int_{0}^{\pi}\int_{-\xi-\pi/2}^{-\xi+\pi/2} I(\phi, \theta) \cos(\phi + \xi) \sin^{2}\theta d\phi d\theta.
\end{equation}
There are no current observations which can constrain the $\theta$-dependence of $I$, but for edge-on orbits the latitudinal dependence of the intensity is unimportant since one can define $J(\phi) = \int_{0}^{\pi} I(\phi,\theta) \sin^{2}\theta d\theta$, which represents the flux contribution from an infinitesimal slice of the planet when viewed face-on. The integrated flux from the planet at a given point in its orbit is then
\begin{equation} \label{edge_on_integral}
F(\xi) = \int_{-\xi-\pi/2}^{-\xi+\pi/2} J(\phi) \cos(\phi + \xi) d\phi.
\end{equation}

It is useful to think of this integral as a convolution $F(\xi) = \int_{0}^{2\pi} J(\phi) K(\phi,\xi) d\phi$, with the piece-wise defined kernel $K$:
\begin{equation}
K(\phi, \xi) = \left\{ \begin{array}{ll}
\cos(\phi + \xi) & \textrm{if~} \cos(\phi + \xi) \ge 0\\
0 & \textrm{if~} \cos(\phi + \xi) < 0.
\end{array} \right.
\end{equation}
The kernel, which represents the response of the phase function to a delta function in $J$, is very broad, with a full width at half-maximum of $2\pi/3$.

\section{From Light Curves to Maps}
In the previous section we developed an analytic expression, Equation~\ref{edge_on_integral}, for the convolution $J(\phi) \to F(\xi)$. There is no closed expression for the deconvolution, $F(\xi) \to J(\phi)$, and there is no guarantee that solutions are unique. Furthermore, two problems arise in practice: the light curve is only sampled at discrete values of $\xi$ which may not span a full planetary rotation, and the measurements of $F(\xi)$ are not arbitrarily precise but instead have associated uncertainties $\sigma_{F}$. Given these realities, it is useful to develop model maps which simplify the integral in Equation~\ref{edge_on_integral} and then use numerical methods to solve the problem or---better yet--- allow for direct inversion.

\subsection{N-Slice (AKA ``Orange Slice'') Model}
The planet is divided into $N$ longitudinal slices of width $\Delta \phi = 2\pi/N$. Each slice has a uniform intensity in both longitude and latitude.  This flux distribution is not ruled out by the observations and---more importantly---smoothing the steps does not significantly change the light curve, provided the total flux from each slice and their brightness-weighted longitude are unchanged. The $j$th slice is centered on $\phi_{j} = \phi_{0} + j\Delta\phi$, where the phase offset, $0 \le \phi_{0} < \Delta \phi$, is useful to accommodate slight discrepancies between the light curve maximum and superior conjunction. The intensity map for the planet is given by:
\begin{equation}
J(\phi) = \left\{ \begin{array}{ll}
J_{0} & \textrm{if $\phi_{0}-\Delta\phi/2 \le \phi < \phi_{0}+\Delta\phi/2$}\\
J_{1}  & \textrm{if $\phi_{1}-\Delta\phi/2 \le  \phi < \phi_{1}+\Delta\phi/2$}\\
\vdots & \vdots\\
J_{N-1}  & \textrm{if $\phi_{N - 1}-\Delta\phi/2 \le  \phi < \phi_{N-1}+\Delta\phi/2$}.
\end{array} \right.
\end{equation}

Since, in practice, one is only ever concerned with comparing the model phase function to data at a finite number of discrete phases, the transformation from N-slice map to light curve can be expressed in matrix form: $F = G J$. The matrix $G$ is defined as $G = \sin\alpha_{+}-\sin\alpha_{-}$, where $\alpha_{+}$ and $\alpha_{-}$ represent the leading and trailing edges of the slices: 
\begin{equation}
\alpha_{\pm} = \cos^{-1}\left(\max[\cos(\xi_{i} + \phi_{j} \pm \Delta\phi/2), 0]\right).
\end{equation}

\subsection{Sinusoidal Model}
\begin{figure}[htb]
\includegraphics[width=84mm]{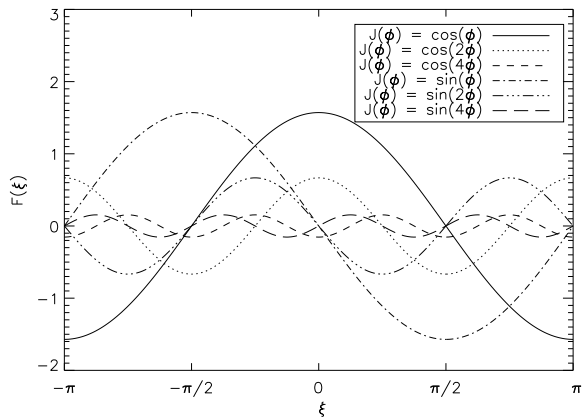}
\caption{The phase function for different sinusoidal maps. Odd sinusoidal maps (starting with $J(\phi)=\cos(3\phi)$) are invisible due to symmetry.}
\label{sinusoidal_kernel}
\end{figure}

Sinusoidal basis maps have the advantage of producing sinusoidal light curves \emph{of the same frequency and phase offset}. If a planet map is composed of sinusoids, $J(\phi) = A_{0}+\sum_{j=1}^{N} A_{j}\cos(j\phi) + B_{j}\sin(j\phi)$, the light curve is simply given by $F(\xi) = F_{0}+\sum_{j=1}^{N} C_{j}\cos(j\xi) + D_{j}\sin(j\xi)$. The coefficients of $F$ are related to those of $J$ by: 
\begin{equation} \label{sinusoidal}
\left(
\begin{array}{c}
F_{0}\\
C_{1}\\
D_{1}\\
\vdots\\
C_{j}\\
D_{j}
\end{array}
\right) = \left(
\begin{array}{c}
2\\
\frac{\pi}{2}\\
-\frac{\pi}{2}\\
\vdots\\
-\frac{2}{j^{2}-1}(-1)^{j/2}\\
\frac{2}{j^{2}-1}(-1)^{j/2}
\end{array}
\right) \left(
\begin{array}{c}
A_{0}\\
A_{1}\\
B_{1}\\
\vdots\\
A_{j}\\
B_{j}
\end{array}
\right),
\end{equation}
where $j$ must be even. Sinusoidal modes with odd $j$ (other than $j=1$) do not have a phase function signature. In Figure~\ref{sinusoidal_kernel} we show the light curve contributions for a handful of sinusoidal modes, assuming that all of the modes have the same amplitude in $J$. The higher frequency modes are strongly suppressed due to the broad smoothing kernel.  This low-pass filter limits the number of modes which can be meaningfully fit with a given light curve. The uncertainty in a sinusoidal mode in the light curve is related to the uncertainty in the map sinusoidal modes by Equation~\ref{sinusoidal} (eg: $\sigma_{C_{1}} = (\pi/2) \sigma_{A_{1}}$).

\subsection{Inversion Techniques}
Since both the N-slice and sinusoidal models described above provide computationally efficient ways to generate light curves from maps, one can use a $\chi^{2}$ fitting routine (Markov Chain Monte Carlo, Levenberg-Marquardt, etc.)  to produce a map from a given light curve. It is simply necessary to demand that $J(\phi)$ be strictly positive. These techniques have the advantage of naturally producing error estimates for the resulting map. Although a unique best-fit map can always be determined in this way, the uncertainty in the fit parameters may be very large if the number of parameters is not commensurate with the signal-to-noise of the light curve.

For the N-slice model, the uncertainty balloons for \emph{all} the $J_{i}$ if too many free parameters are used. We therefore suggest running multiple fits with different numbers of slices. When the addition of a slice does not improve the $\chi^{2}$, one has achieved the best model that the data can support. Instead of repeating the fit with fewer slices, one can apply smoothing to the map in the form of a Bayesian prior \citep[eg:][]{Knutson_2007}. For well-designed priors, however, there is no fundamental difference between models with large $N$ and long smoothing lengths, versus models with smaller $N$ and little or no smoothing. In the interest of simplicity we recommend using fewer slices rather than smoothing.

An observed light curve can be quickly deconvolved into sinusoidal maps by determining its Fourier components, $F_{0}$, $C$ and $D$, then converting them via the Equation~\ref{sinusoidal}. It is expedient to assume that there is no power in the odd modes (other than $j=1$) to avoid degenerate solutions, but this may lead to a systematic error in the model map, depending on how much power is present in these modes in the real map. The uncertainty in the map parameters may be determined using a $\chi^{2}$ routine or Monte Carlo analysis. Since the sinusoidal model has linearly independent modes, only the uncertainty in the highest-frequency modes explodes when too many modes are considered. As a rule of thumb, one should truncate the Fourier series once the uncertainty in coefficients becomes greater than the coefficients themselves.    
  
\subsection{Inversion Example}
\begin{figure}[htb]
\includegraphics[width=84mm]{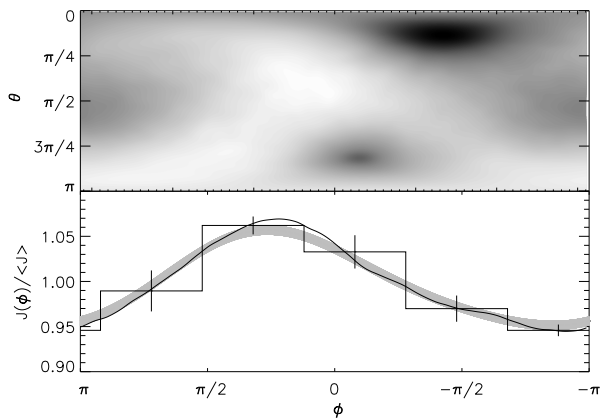}
\protect\caption{The top panel shows a brightness map (darker is cooler) based on the atmospheric dynamics model of \cite{Cho_2003}, with $T_{\rm min}=1020$~K and $T_{\rm max}=1269$~K. In the lower panel, the solid line is $J(\phi)$, the histogram and associated error bars represents the best-fit 5-slice model, and the gray band is the $1\sigma$ confidence interval for the sinusoidal map. Note that positive $\phi$ are to the left of the plot, to facilitate comparison with the light curve in Figure~\ref{lc_cho_03}.}
\label{map_cho_03}
\end{figure}

\begin{figure}[htb]
\includegraphics[width=84mm]{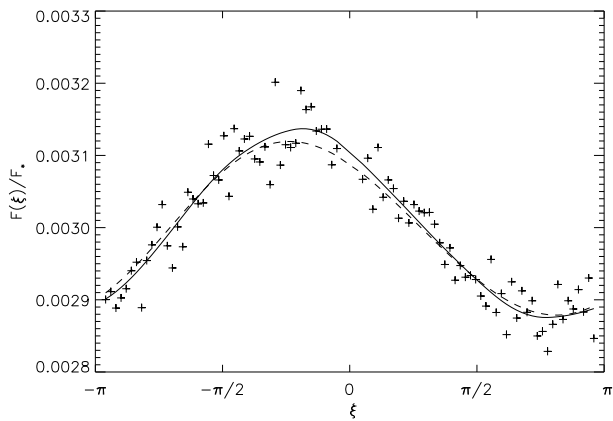}
\protect\caption{The phase function light curve for the planet map shown in Figure~\ref{map_cho_03}, convolved with photometric scatter comparable to \cite{Knutson_2007}. The solid and dashed lines shows the light curves of the best-fit N-slice and sinusoidal models, respectively.}
\label{lc_cho_03}
\end{figure}

We now turn to an example map and test the ability of the algorithms described above to recover the correct features of this map. The top panel of Figure~\ref{map_cho_03} shows the brightness map computed from a snapshot of the atmospheric dynamics model of \cite{Cho_2003}. We performed an analogous test---and obtained comparable results---using a snapshot from the model of \cite{Showman_2008}, which we do not include here in the interest of space. The brightness map was generated by treating each pixel of the temperature map as a blackbody and computing an associated intensity at $8 \mu$m. The bottom panel of Figure~\ref{map_cho_03} shows the integrated longitudinal brightness map, $J(\phi)$. Note that the $\sin^{2}\theta$ term in the integral for $J$ attenuates the flux contribution from the poles of the planet. Also shown in the bottom panel are the best fit maps for the N-slice and sinusoidal models. The light curve associated with the map, as well as the best-fit light curves for the models, are shown in Figure~\ref{lc_cho_03}. The map $J$ was converted to an idealized light curve, and mock observations (comprised of 100 data points) were generated by removing the segments of the light curve corresponding to the transit and secondary eclipse, then scaling the planet/star flux ratio and the photometric uncertainties to roughly match those of \cite{Knutson_2007}. Both models reproduce the features of the map, as well as the light-curve. The best-fit 5-slice model was determined using a Levenberg-Marquardt $\chi^{2}$-minimization routine. The sinusoidal map was determined by decomposing the light curve into sinusoidal components to $j=2$, then converting the coefficients using Equation~\ref{sinusoidal}. The uncertainties---estimated by Monte Carlo analysis---in the $j=4$ terms are larger than their amplitudes and are therefore ignored. The insensitivity of phase functions to the first and most important odd mode, $j=3$, leads to $\sim 10$\% errors in the resulting map, based on the maps of \cite{Cho_2003} and \cite{Showman_2008}. We model the effect of limb darkening by adding $[0.8 + 0.2\sin\theta]$ to the expression for $J(\phi)$ and adding $[0.8 + 0.2\cos(\phi + \xi)]$ to the integrand of Equation~\ref{edge_on_integral}. We find the resulting light curve to differ by less than $0.2$\%, justifying our decision to neglect limb darkening in the formalism above. 

\section{Discussion \& Conclusions}
The best current light curves can be represented by the function $F(\xi) = F_{0}+ C_{1}\cos\xi + D_{1}\sin\xi + C_{2}\cos(2\xi) + D_{2}\sin(2\xi)$ which can be directly translated into a longitudinal brightness map using Equation~\ref{sinusoidal}. The $j=3$ modes should cancel out by symmetry so their presence in the light curve would indicate systematic errors; $j\ge4$ modes are generally lost in the noise. The longitudes and amplitudes of the primary hot-spots and cold-spots can be determined from the $j=1$ terms, while the relative strength of the $j=1$ and $j=2$ modes indicates whether there are secondary local maxima/minima. By the same token, a 4-slice model with variable phase offset should be sufficient to model most phase function light curves. For light curves with incomplete phase coverage, the uncertainty in the sinusoidal map is the same at all longitudes whereas the N-slice model naturally has larger uncertainties for slices which were visible for less time, an intuitive and desirable property. So far only half-orbits have been allocated to phase function studies, but the warm Spitzer mission will provide a perfect opportunity to obtain light curves spanning full planetary orbits \citep{Deming_2007}.

The contribution of sinusoidal modes to the observed light curve decreases precipitously with $j$. Not only does the $J(\phi) \to F(\xi)$ transformation suppress the terms as $1/(j^{2} -1)$, but the intrinsic power of these modes in the underlying map might drops as $\sim j^{-2}$ \citep[as is the case with themap from][]{Cho_2003}. As a result, the $j=4$ modes in light curves will be $\sim 20$ times weaker than the $j=2$ modes and will likely remain undetectable with JWST. Even when higher quality light curves are eventually obtained---and assuming that stellar variability is not the limiting factor---the physical significance of the model map would be questionable due to the insensitivity of the phase function to odd sinusoidal modes. The power in these modes can be constrained theoretically by dynamical atmospheric models and observationally through secondary eclipse mapping, which promises to be feasible with JWST. 

\acknowledgments
N.B.C. is supported by the Natural Sciences and Engineering Research Council of Canada. E.A. is supported by a National Science Foundation Career Grant. Support for this work was provided by NASA through an award issued by JPL/Caltech. The authors wish to thank E. Rauscher and A. Showman for use of their model temperature maps.

%\clearpage

\end{document}